 \documentclass[aps,pra,twocolumn,groupedaddress,amsmath,amssymb,showpacs]{revtex4}
\usepackage{graphicx}
\usepackage{amsmath}
\usepackage{amssymb}
\usepackage{mathrsfs}
\usepackage{amsfonts}
\usepackage{dcolumn}
\usepackage{bm}

\providecommand{\abs}[1]{\lvert#1\rvert}

\begin{document}
\title{How orbital angular momentum affects beam shifts in optical reflection}
\author{M.~Merano}
\affiliation{Huygens Laboratory, Leiden University,
P.O.\ Box 9504, 2300 RA Leiden, The Netherlands}
\author{N. Hermosa}
\affiliation{Huygens Laboratory, Leiden University,
P.O.\ Box 9504, 2300 RA Leiden, The Netherlands}
\author{A.~Aiello}
\affiliation{Max Planck Institute for the Science of Light, G\"{u}nter-Scharowsky-Stra{\ss}e 1/Bau 24, 91058 Erlangen, Germany}
\author{J.~P.~Woerdman}
\affiliation{Huygens Laboratory, Leiden University,
P.O.\ Box 9504, 2300 RA Leiden, The Netherlands}
\begin{abstract}
It is well known that reflection of a Gaussian light beam ($\text{TEM}_{00}$) by a planar dielectric interface leads to four beam shifts when compared to the geometrical-optics prediction. These are the spatial Goos-H\"{a}nchen (GH) shift, the angular GH shift, the spatial Imbert-Fedorov (IF) shift and the angular IF shift. We report here, theoretically and experimentally,  that endowing the beam with Orbital Angular Momentum (OAM) leads to coupling of these four shifts; this is described by a $4 \times 4$ mixing matrix.
\end{abstract}
\pacs{42.79.-e, 41.20.Jb, 42.25.Gy, 78.20.-e } \maketitle
\section{Introduction}
 The reflection of a light beam by a mirror shows subtle aspects that were first conjectured by Newton \cite{NewtonOptiks}: the center of the reflected beam may show a small spatial shift \emph{in} the plane of incidence, relative to the position predicted by geometrical optics. This shift has been named after Goos and H\"{a}nchen (GH) who were the first to observe it in total internal reflection (TIR) \cite{GH}.   Additionally, there is a spatial shift \emph{perpendicular} to the plane of incidence, the so-called Imbert-Fedorov shift (IF) \cite{Imbert,Fedorov}. There exist also \emph{angular} GH and IF shifts, both of which have been demonstrated recently in external reflection \cite{NatPhoton.3.337,HostenandKwiat}. The angular shifts can be seen as shifts in wave-vector space \cite{HostenandKwiat,BliokhNP,AielloOL08}. All these shifts depend on the polarization of the incident photons. Accurate calculations of either  GH or IF shifts (or both) can be found in Refs. \cite{Artmann,BliokhPRL,BliokhPRE}. In more recent years the GH shift has been studied in a large diversity of cases, ranging from photonic crystals \cite{Felbacq} to neutron optics \cite{Haan}.

We are interested in the question how these beam shifts are affected when the light beam is endowed with Orbital Angular Momentum (OAM). OAM is a relatively novel degree of freedom of a light beam that has found applications from optical tweezers to quantum information science \cite{Mair,He}. Theoretically, a  treatment of the effect of OAM on beam shifts has aready been given, first by Fedosoyev \cite{Fedoseyev01,Fedoseyev08,Fedoseyev20091247} and then by Bliokh \textit{et al.}  \cite{Bliokh:09}. Here we prefer to develop our own theoretical treatment based on straightforward  application of Snell's law and the Fresnel equations, in order to derive a unified matrix formalism for the four basic shifts: spatial GH, angular GH, spatial IF and angular IF. Experimentally, Okuda and Sasada have studied the deformation  of an OAM carrying beam by TIR very close to the critical angle \cite{Okuda}; however, they did not report GH and IF shifts. Dasgupta and Gupta have measured the IF shift of an OAM beam reflected by a dielectric interface, but only for the spatial case \cite{Dasgupta}.

It is the purpose of this article to report a theoretical and experimental study of the effect of OAM on the four basic shifts: spatial GH, angular GH, spatial IF and angular IF. We find that these shifts are coupled by OAM; this is described by an OAM dependent $4 \times 4$ mixing matrix. We have experimentally confirmed this mixed occurrence of GH and IF shifts.
\section{Theory}
In this section we furnish a thorough theoretical analysis for the problem of the reflection of an OAM carrying light beam by a dielectric interface.

Consider a monochromatic beam containing a continuous distribution of wave-vectors $\mathbf{k}$ centered around  $\mathbf{k}_0 = k_0 \hat{\bm{z}}_i$, where $\hat{\bm{z}}_i$ is a unit vector along the central propagation direction of the incident beam: $\mathbf{k} = k_0 \hat{\mathbf{k}} = \mathbf{k}_0 + \mathbf{q}$, with
$\mathbf{q} = \mathbf{q}_T + q_L \hat{\bm{z}}_i$ and $\mathbf{q}_T \cdot \hat{\bm{z}}_i = 0$. Using the notation of Fig. 1, we write $ q_T/k_0 = \sin \alpha$ and $q_L/k_0 = 1- \cos \alpha$ with $q_T=\abs{\mathbf{q}_T}$ and $\alpha = \arccos (\hat{\mathbf{k}} \cdot \hat{\mathbf{k}}_0)$.
%
%
\begin{figure}[!b]
\includegraphics[angle=0,width=7.5truecm]{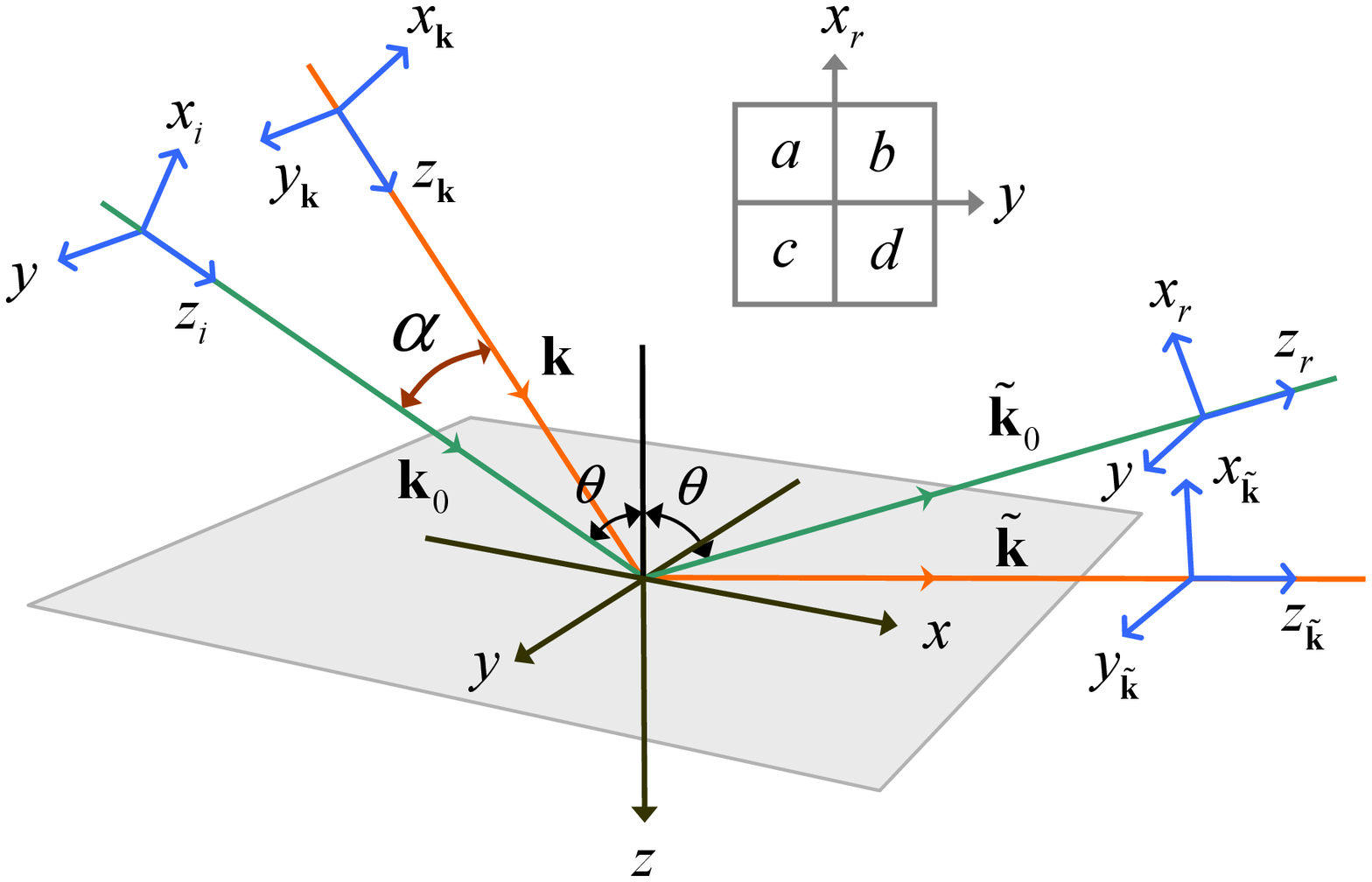}
\caption{\label{fig:1} (Color online) Geometry of beam reflection at a
dielectric interface. The reflected wavevector $\tilde{\mathbf{k}}$ is the mirror image of the incident wavevector $\mathbf{k}$. Inset shows quadrant detector with sensitive areas $a,b,c,d$.
}
\end{figure}
%
%
%
A collimated beam has a narrow distribution of wave-vectors around  $\mathbf{k}_0$ such that $\sin \alpha \cong \alpha \ll 1$ with $ q_T/k_0 \cong \alpha \ll 1$ and $q_L/k_0 \cong (q_T/k_0)^2/2$. Thus, if we write $\mathbf{k} = k_0 (\hat{\bm{x}}_i U +\hat{\bm{y}} V +\hat{\bm{z}}_i W )$ with $W=\sqrt{1 - U^2 - V^2} $, we can assume $U,V \ll 1$ without significant error.
Let $\mathbf{E}^I(\mathbf{r},t)$ be the electric field of the incident beam. Upon reflection this field evolves to $\mathbf{E}^R(\mathbf{r},t)$ which is to be found. From the linearity of the wave equation it follows that  $\mathbf{E}^R(\mathbf{r},t)$ can be determined by studying the action of the interface upon each plane wave field
\begin{align}\label{th10}
\bm{A}^I(\mathbf{k}) = \bm{f}_\perp(\mathbf{k} ) \exp{(i \mathbf{k}} \cdot \mathbf{r} - i\omega t)
\end{align}
that constitutes $\mathbf{E}^I(\mathbf{r},t)$, with $\omega = \abs{\mathbf{k}} c$.
According to Refs. \cite{FainmanANDShamir,Aiello:09}, we assume the polarization dependent amplitude of $\bm{A}^I(\mathbf{k})$ equal to  $\bm{f}_\perp(\mathbf{k} ) = \hat{\bm{f}} - \hat{\mathbf{k}} ( \hat{\mathbf{k}} \cdot \hat{\bm{f}} ) = a_p(\mathbf{k}) \; \hat{\bm{x}}_\mathbf{k}  + a_s(\mathbf{k}) \; \hat{\bm{y}}_\mathbf{k} $ with
\begin{align}\label{th20}
a_p(\mathbf{k}) = f_p + V f_s \cot \theta, \qquad a_s(\mathbf{k}) = f_s - V f_p \cot \theta,
\end{align}
up to first order in $U,V$, and $\theta = \arccos (\hat{\bm{z}}_i \cdot \hat{\bm{z}})$.
Here
$ \hat{\bm{f}}  = f_p \hat{\bm{x}}_i + f_s \hat{\bm{y}}$ is a unit complex vector that fixes the polarization of the incident beam, and
  $\hat{\bm{y}}_\mathbf{v} = \hat{\bm{z}} \times \mathbf{v} / \abs{\hat{\bm{z}} \times \mathbf{v}} $,  $\hat{\bm{x}}_\mathbf{v} = \hat{\bm{y}}_\mathbf{v}\times \mathbf{v}$ denote a pair of mutually orthogonal real unit vectors that together with the arbitrary vector $\hat{\mathbf{v}}=\mathbf{v}/ \abs{\mathbf{v}}$ form a right-handed Cartesian reference frame $K_\mathbf{v} = \{ \hat{\bm{x}}_\mathbf{v},\hat{\bm{y}}_\mathbf{v},\hat{\mathbf{v}}\}$ attached to $\mathbf{v}$.

When the beam is reflected at the interface, each plane wave evolves as: $ \bm{A}^I(\mathbf{k}) \rightarrow \bm{A}^R(\mathbf{k})$ where
\begin{align}\label{th30}
\bm{A}^R(\mathbf{k})=\left[ r_p(\theta_\mathbf{k})  \, a_p  \, \hat{\bm{x}}_{\tilde{\mathbf{k}}}  + r_s(\theta_\mathbf{k})  \, a_s \, \hat{\bm{y}}_{\tilde{\mathbf{k}}} \right] \chi(\tilde{\mathbf{r}},t),
\end{align}
and $\chi(\tilde{\mathbf{r}},t) = \exp{(i \tilde{\mathbf{k}} \cdot \mathbf{r} - i\omega t)}=\exp{(i {\mathbf{k}} \cdot \tilde{\mathbf{r}} - i\omega t)}$. The notation $\tilde{\mathbf{v}}$ indicates the mirror image of the vector $\mathbf{v}$ with respect to the interface:
 $\tilde{\mathbf{v}} = \mathbf{v} - 2 \hat{\bm{z}} (\hat{\bm{z}} \cdot \mathbf{v})$, with $\tilde{\mathbf{v}} \cdot \mathbf{u} = \mathbf{v} \cdot \tilde{\mathbf{u}}$ \cite{Gragg}. Moreover, $r_p(\theta_\mathbf{k})$ and $r_s(\theta_\mathbf{k})$ are the Fresnel reflection coefficients at incidence angle $\theta_\mathbf{k} = \arccos(\hat{\mathbf{k}} \cdot \hat{\bm{z}})$ for $p$ and $s$ waves, respectively. By direct calculation it is not difficult to show that, up to first order in $U,V$,
\begin{align}
\hat{\bm{x}}_{\tilde{\mathbf{k}}} = & \; \hat{\bm{x}}_r - V \cot \theta \hat{\bm{y}} + U \hat{\bm{z}}_r, \label{th40a} \\
\hat{\bm{y}}_{\tilde{\mathbf{k}}} = & \; V \cot \theta \hat{\bm{x}}_r  + \hat{\bm{y}} - V \hat{\bm{z}}_r, \label{th40b} \\
r_\lambda(\theta_{\mathbf{k}})= & \; r_\lambda + U  r_\lambda', \label{th50}
\end{align}
%
%
%
%
%
where $ \lambda \in \{p, s \}$, $r_\lambda = r_\lambda(\theta)$, and $r_\lambda' = \partial r_\lambda(\theta)/ \partial \theta$.
With the use of Eqs. (\ref{th20},\ref{th40a}-\ref{th50}) into Eq. (\ref{th30}) we obtain
\begin{align}
\bm{A}^R(\mathbf{k})= \hat{\bm{x}}_r A_p^R(\mathbf{k}) + \hat{\bm{y}} A_s^R(\mathbf{k}) + \hat{\bm{z}}_r A_L^R(\mathbf{k}),
\end{align}
 where, up to first order in $U,V$,
\begin{align}
A_\lambda^R(\mathbf{k}) = & \;  f_\lambda r_\lambda \left( 1 + i X_\lambda U - i Y_\lambda V \right) \chi(\tilde{\mathbf{r}},t),   \label{th65a} \\
A_L^R(\mathbf{k}) =   & \; \left( f_p r_p  U - f_s r_s  V\right) \chi(\tilde{\mathbf{r}},t). \label{th65c}
\end{align}
Here, we have defined
\begin{align}\label{th70}
X_p = -i \frac{\partial \ln r_p}{\partial \theta}, \qquad Y_p = i \frac{f_s}{f_p}\left( 1 + \frac{r_s}{r_p}\right) \cot \theta,
\end{align}
with  $X_s =  \left. X_p \right|_{p \leftrightarrow s}$ and $Y_s =  \left. - Y_p  \right|_{p \leftrightarrow s}$. The limit of specular reflection is achieved by letting $r_p \rightarrow 1$ and $r_s \rightarrow -1$ where Eq. (\ref{th70}) reduces to $X_p = 0 =X_s$ and $Y_p = 0 =Y_s$.
Notice that from Eqs. (\ref{th65a}-\ref{th65c}) it follows that for a paraxial beam the longitudinal electric field energy density $\abs{A_L^R}^2$  scales as $\sim \alpha^2$ and it is therefore negligible with respect to the transverse electric field energy density $\abs{A_p^R}^2 + \abs{A_s^R}^2$  that scales as $\sim 1 +  2 \alpha$. Thus, up to first order in $\alpha$, we can neglect the longitudinal term $A_L^R(\mathbf{k})$ and write $\bm{A}^R(\mathbf{k}) \simeq \hat{\bm{x}}_r A_p^R + \hat{\bm{y}} A_s^R $.
Moreover,  for small shifts $X_\lambda$ and $Y_\lambda$ one can write $1 + i X_\lambda U - i Y_\lambda V \simeq \exp(i X_\lambda U - i Y_\lambda V)$,  and in the Cartesian coordinate system attached to the reflected beam  $ \chi(\tilde{\mathbf{r}},t) = \exp[i(- U X + V Y + W Z)] \exp(-i \omega t)$,  with $X = k_0 x_r, \; Y = k_0 y$, and $ Z = k_0 z_r$, where $z_r$ is the distance from the waist of the incident beam to the quadrant detector measured along the trajectory of the beam. Thus, Eq. (\ref{th65a}) can be rewritten as:
\begin{align}
A_\lambda^R(\mathbf{k}) \simeq & \;  f_\lambda r_\lambda \chi(-X +X_\lambda,Y - Y_\lambda,Z ,t).  \label{th80a}
\end{align}

The passage from the single plane wave field $\bm{A}^R(\mathbf{k})$ to the total electric field $\mathbf{E}^R(\mathbf{r}, t)$ is realized by substituting the plane wave scalar amplitude   $\chi({\mathbf{r}},t)$ into Eq. (\ref{th80a}), with the electric field scalar amplitude $E({\mathbf{r}},t)$ describing the spatial distribution of the incident beam.
 In the present case, as we want to study the behavior under reflection of OAM beams,
we choose $E({\mathbf{r}},t) = \psi_\ell (\mathbf{r}) \exp(- i \omega t)$, being $ \psi_\ell (\mathbf{r})$ the Laguerre-Gauss paraxial field with OAM index $\ell \in \{0, \pm 1, \pm 2,\ldots \}$ and radial index $p=0$: $ \psi_\ell (X,Y,Z) \propto \exp [ -{(X^2 + Y^2)}/ (2 \Lambda + i 2 Z)]\left(X + i s_\ell Y \right)^{\abs{\ell}}$,
with $s_\ell = \mathrm{sign} (\ell)$ and $\Lambda = k_0(k_0 w_0^2/2)$ denoting the dimensionless Rayleigh range of the beam with waist $w_0$ \cite{MandelBook}. Thus, the transverse electric field  of a Laguerre-Gauss beam reflected by a plane interface can be written as:
\begin{align}\label{th90}
E_\lambda^R (\mathbf{r},t) \simeq    f_\lambda  r_\lambda   \psi_\ell(-X +X_\lambda,Y - Y_\lambda,Z )\exp(- i \omega t).
\end{align}
In this expression the terms $X_\lambda$ and $Y_\lambda$ are responsible for the
GH \cite{AielloOL08} and IF \cite{Bliokh:09} shifts of the center of the beam, respectively.
These displacements can be assessed by measuring the position of the center of the reflected beam with a quadrant detector centered at $x_r=0, \, y=0$ along the reference axis $z_r$ attached to the reflected central wave vector $\widetilde{\mathbf{k}}_0 = k_0 \hat{\mathbf{z}}_r$. A quadrant detector has four
sensitive areas, denoted with $a,b,c,d$ in the inset of Fig. 1, each delivering a photocurrent $I_a,I_b,I_c,I_d$ respectively, when illuminated. The two currents $I_x = (I_a + I_b) - (I_c + I_d)$ and $I_y = (I_b + I_d) - (I_a + I_c)$ are thus proportional to the $x$- and the  $y$-displacement of the beam intensity distribution relative to the center  of the detector, respectively.

If $\ell =0$, $\psi_0(-X + X_\lambda, Y - Y_\lambda)$ reduces to a shifted fundamental Gaussian beam, and in the hypothesis of small deviations $X_\lambda, Y_\lambda \ll 1$, a straightforward calculation  furnishes
\begin{align}
\frac{I_x}{I} = N_0 \left( \Delta_\text{GH} + \frac{Z}{\Lambda} \Theta_\text{GH}  \right), \quad
\frac{I_y}{I} = N_0 \left( \Delta_\text{IF} + \frac{Z}{\Lambda} \Theta_\text{IF}  \right), \label{med1}
\end{align}
where $I =I_a + I_b +I_c + I_d$, and $ N_0 = \sqrt{2/(\pi \sigma^2)}$ with $\sigma^2 = (\Lambda /2)\sqrt{1 + Z^2/\Lambda^2}$.
Here we have defined the two spatial ($\Delta$)  and the two angular ($\Theta$) shifts
\begin{align}
\Delta_\text{GH} = \sum_{\lambda =p,s}  w_\lambda \text{Re}(X_\lambda), & \; & \Delta_\text{IF} = \sum_{\lambda =p,s} w_\lambda \text{Re}(Y_\lambda), \label{spatial}
\end{align}
and
\begin{align}
\Theta_\text{GH} = \sum_{\lambda =p,s}  w_\lambda \text{Im}(X_\lambda),& \;  & \Theta_\text{IF} = \sum_{\lambda =p,s}  w_\lambda \text{Im}(Y_\lambda), \label{angular}
\end{align}
respectively, where the nonnegative coefficients $w_\lambda$ are defined as the fraction of the electric field energy with polarization $\lambda=p,s$ in the reflected beam:
\begin{align}
w_\lambda \equiv & \; \frac{\abs{r_\lambda f_\lambda}^2}{\abs{r_p f_p}^2 + \abs{r_s f_s}^2} \label{th120a}.
\end{align}

If $\ell \neq 0$, Eq. (\ref{med1}) becomes
\begin{align}
\frac{I_x}{I} = N_\ell \left(\Delta_\text{GH}^\ell + \frac{Z}{\Lambda} \Theta_\text{GH}^\ell \right), \quad
\frac{I_y}{I} = N_\ell \left( \Delta_\text{IF}^\ell + \frac{Z}{\Lambda} \Theta_\text{IF}^\ell  \right), \label{med2}
\end{align}
where $N_\ell = N_0 {\Gamma(\abs{l} + 1/2)}/\left[ \Gamma(\abs{l} + 1) \sqrt{\pi} \right]$ ($\Gamma(x)$ denotes the Gamma function), and
\begin{align}
\left[
  \begin{array}{l}
    \Delta_\text{GH}^\ell \\
    \Theta_\text{IF}^\ell  \\
    \Delta_\text{IF}^\ell  \\
    \Theta_\text{GH}^\ell  \\
  \end{array}
\right]  =   \left[
             \begin{array}{cccc}
                1  &  -2 \ell  &  0  & 0 \\
               0 & 1 +\abs{2 \ell} & 0 & 0 \\
               0 & 0 & 1 & 2 \ell \\
               0 & 0 & 0 & 1+\abs{2 \ell} \\
             \end{array}
           \right]
 \left[
  \begin{array}{l}
    \Delta_\text{GH} \\
    \Theta_\text{IF} \\
    \Delta_\text{IF} \\
    \Theta_\text{GH} \\
  \end{array}
\right]. \label{mixing}
\end{align}
Equation (\ref{mixing}) clearly displays the mixing between spatial and angular GH and IF shift, occurring only for $\ell \neq 0$, and it is in agreement with the results presented in Ref. \cite{Bliokh:09}, apart from the factor ``$2$'' in front of $\ell$ \cite{Note1}. Notice that the polarization dependence of the four $\ell$-dependent shifts on the left side of Eq. (\ref{mixing}) resides in the four $\ell$-independent shifts on the right side of the same equation. It turns out that the $4 \times 4$ mixing matrix itself is  polarization-independent.
It should be noticed that in TIR, contrary to partial reflection, both GH and IF angular shifts $ \Theta_\text{GH}$ and $ \Theta_\text{IF}$ are identically zero since the Fresnel coefficients are purely imaginary \cite{AielloPra09}. Thus, in this case it follows from Eq. (\ref{mixing}) that mixing vanishes.
\section{Experimental set-up}
Our experimental set-up is shown in Fig. 2.
%
%
\begin{figure}[!h]
\includegraphics[angle=0,width=8truecm]{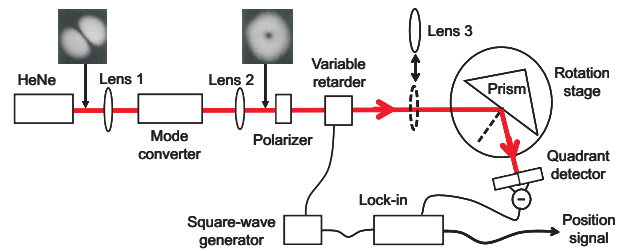}
\caption{\label{fig:2} (Color online) Experimental set-up. The insets show the $\mathrm{HG}_{10}$ and $\mathrm{LG}_{10}$ mode profiles. The quadrant detector measures the OAM controlled shift of the reflected beam in the plane of incidence (GH shift) and perpendicular to it (IF shift). Both the GH and the IF shift have a spatial and an angular contribution. See text for further details.
}
\end{figure}
%
%
A home built HeNe laser ($\lambda_0 = 633 \, \text{nm}$) is forced to operate in a single higher-order Hermite-Gaussian ($\text{HG}_{nm}$) mode with $m=0$ by insertion of a $40\text{-}\mu \text{m}$ diameter wire normal to the axis of the laser cavity \cite{Beijersbergen93}. The $\text{HG}_{n0}$ beam is sent through an astigmatic mode converter consisting of  two cylinder lenses, with their common axis oriented at $45^\circ$ relative to the intra-cavity wire. This introduces a Gouy phase which converts the $\text{HG}_{n0}$ beam in a $\text{LG}_{\ell p}$ beam with $\ell = n$ and $p = 0$ \cite{Beijersbergen93}. Lenses $1$ and $2$ are used for mode matching; the beam leaving lens $2$ is collimated with a waist parameter $w_0=775 \; \mu \text{m}$, a power of  typically $600 \;\mu \text{W}$ and a polarization set by a linear polarizer. We have incorporated the option to greatly enhance the angular spread of the beam by inserting lens $3$ ($f=70 \; \mathrm{mm}$), leading to $w_0= 19 \; \mu\text{m}$. Either with or without lens $3$ present, the beam is externally reflected by the base plane of a glass prism (BK7, $n=1.51$).   We measured the polarization-differential shifts of the reflected $\text{LG}_{\ell 0}$ beam with a calibrated quadrant detector. We also obtained these shifts for the fundamental $\text{LG}_{00}$ beam ($=\text{TEM}_{00}$) by simply removing the intra-cavity wire from the HeNe laser.

It follows from Eqs. (\ref{med1}-\ref{med2}) that using a collimated incident beam, i.e. $\Lambda \gg Z$, leads to total predominance  of the spatial shift. On the other hand, the use of a focused beam, i.e. $\Lambda \ll Z$,  leads to total predominance of the angular shift.  These two extreme cases were realized in our experiment by removal respectively insertion of lens $3$. Specifically, the value of the Rayleigh range $L = k_0 w_0^2/2$ was $2.96 \; \text{m}$ and $1.8 \; \text{mm}$, respectively; as standard we have chosen the distance $z_r$ between the beam waist and the quadrant detector to be $9.5 \; \text{cm}$. We experimentally checked the angular nature of the shift (where expected) by verifying that the detector signal depended linearly on changes in $z_r$.

We performed all measurements by periodically ($2.5 \; \text{Hz}$) switching the polarization of the incident beam with a liquid-crystal variable retarder and by synchronously measuring (with a lock-in amplifier) the relative beam position for one polarization with respect to the other \cite{NatPhoton.3.337,Merano07}. Experimentally we were restricted to using the
first-order $\text{LG}$ modes ($\ell = \pm 1$) by the low gain of the HeNe laser.
\section{Experimental  results and comparison with theory}
Our experimental results for the polarization-differential shifts versus the angle of incidence are reported in Fig. 3, together with the theoretical curves ($\ell =0$ and $\ell =\pm 1$) which are based upon Eqs. (\ref{med1}-\ref{med2}).
%
%
\begin{figure}[!h]
\includegraphics[angle=0,width=8.5truecm]{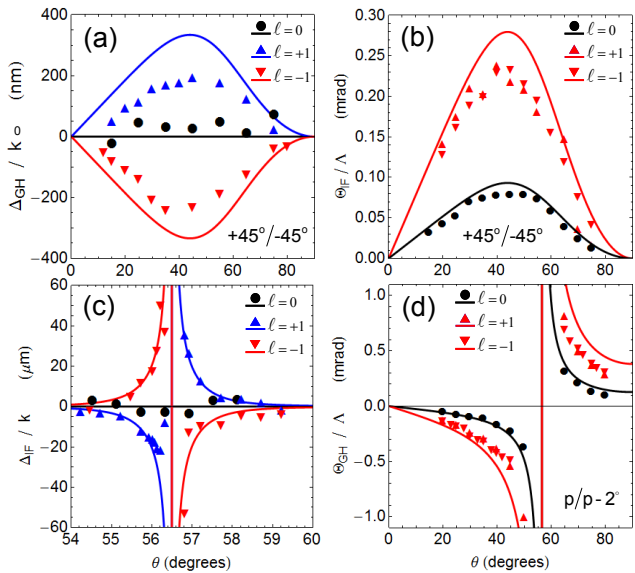}
\caption{\label{fig:3} (Color online)  Reflective beam shift for partial dielectric reflection from an air-glass interface as a function of the angle of incidence. Plotted curves are the theoretical polarization-differential shifts for the two polarizations indicated in each panel. Experimental data and theoretical curves refer to $\ell =0$ and $\ell =\pm 1$. The panels display the spatial GH shift (a), angular IF shift (b), spatial IF shift (c) and angular GH shift  (d). Here $k_0 = 2 \pi /\lambda_0$; see text for further details.
}
\end{figure}
%
%
The four panels show the spatial and angular varieties of GH and IF shifts.
 Note that we have plotted here the true GH and IF shifts $\Delta/k_0$ and $\Theta/\Lambda$, respectively, and not the dimensionless shifts $\Delta$ and $\Theta$.
In each individual case the polarization modulation basis has been chosen such as to maximize the magnitude of the OAM effect.
 Fig. 4 shows a cartoon-like representation of the four cases that we address.
 %
 %
%
\begin{figure}[!h]
\includegraphics[angle=0,width=8.5truecm]{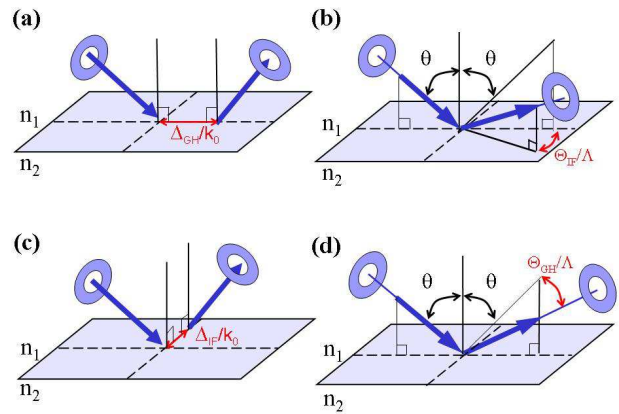}
\caption{\label{fig:4} (Color online)  Reflection of an orbital-angular-momentum (OAM) carrying Laguerre-Gaussian (LG) beam at a dielectric interface. Depending on the input polarization this may lead to a longitudinal shift (Goos-H\"{a}nchen effect) or to a transverse shift (Imbert-Fedorov effect), where longitudinal and transverse refer to the plane of incidence. Each of these shifts consists of a spatial part and an angular part, which are observed, respectively, in the near field and in the far field of the LG beam. The magnitude of the shifts increases with the OAM index $\ell$. Panel (a) shows the spatial Goos-H\"{a}nchen effect, panel (b) the angular Imbert-Fedorov effect, panel (c) the spatial Imbert-Fedorov effect and panel (d) the angular Goos-H\"{a}nchen effect. Note that $\Delta$ and $\Theta$ are dimensionless quantities; $\Lambda$ and $k_0$ are the dimensionless Rayleigh range and the wavenumber of the LG beam, respectively. See text for further details.
}
\end{figure}
%
%
 The overall agreement between experiment and theory is reasonable if we realize that there is no fitting parameter involved; we ascribe the remaining discrepancies to insufficient modal purity of the $\text{LG}_{10}$ beam (we are very sensitive to this since we use a quadrant detector).

Fig. 3a shows the spatial GH shift for a polarization basis of diagonal linear polarizations. In this case, the GH shift is absent for $\ell =0$ but it appears for $\ell =\pm 1$; the sign of the shift reverses when going from  $\ell = +1$ to $\ell = -1$. In Fig. 3b we show that the angular IF shift is different for $\ell =0$ and $\ell =\pm 1$, using again diagonal linear polarizations. No difference occurs for $\ell =+1$ versus $\ell =-1$. Proceeding to Fig.3c we observe an angular GH shift when using a linear polarization basis $(s,p)$,  for both $\ell =0$ and $\ell =\pm 1$. Both cases show a dispersive resonance at the Brewster angle; for $\ell =0$  these experimental results have been reported recently \cite{NatPhoton.3.337} whereas the data for $\ell =\pm 1$ (with opposite sign for $\ell = +1$ and $\ell = -1$) are new. Fig. 3d shows the OAM dependence of the spatial IF shift, observed in a linear polarization basis $(p,p - 2^\circ)$ \cite{Dasgupta,Note2}. Here the shift is zero for $\ell = 0$ whereas it shows a dispersive Brewster resonance for $\ell =\pm 1$ (with opposite sign for $\ell = +1$ and $\ell = -1$).

Finally, we have confirmed experimentally that OAM did not affect spatial and angular GH and IF shifts in the TIR case (not shown); TIR was realized by flipping the glass prism in Fig. 2.
\section{Conclusions}
We have presented a unified theoretical description of how the Orbital Angular Momentum (OAM) of a light beam affects its kinematic degrees of freedom when the beam is reflected by a dielectric interface. Without OAM the reflection leads to four beam shifts relative to geometrical optics, namely the Goos-H\"{a}nchen (GH) and Imbert-Fedorov (IF) shifts, each of which may have a positional and an angular part. We introduce a $4 \times 4$ polarization-independent (but $\ell$-dependent) coupling matrix that describes the OAM induced mixing of these four shifts when using a quadrant detector. Experimentally, we have confirmed this theory by measuring the four shifts as a function of the angle of incidence, for OAM values $\ell =0$ and $\pm 1$. 
We have observed for the first time the OAM induced spatial GH shift as well as the OAM affected angular GH and IF shifts (see Fig. 3 a, b, c).
Extension of all this from reflection to transmission (i.e. refraction) is straightforward.

Understanding these effects is important since they generally affect control of OAM beams by mirrors and lenses. The angular shifts are particularly interesting from a metrology point of view, both classically and quantum mechanically, since the corresponding transverse excursion of the beam center grows without limits when the beam propagates; this greatly promotes its detectability \cite{NatPhoton.3.337,TrepsPointer}.
\section{Acknowledgments}
%
%
We thank K.Y. Bliokh for stimulating us to measure the OAM-induced GH shift which acted as the seed of this paper. Our work is part of the program of the Foundation for Fundamental Research of Matter (FOM). It is also supported by the European Union within FET Open- FP7 ICT as part  of the STREP program 255914 PHORBITECH. AA acknowledges support from the Alexander von Humboldt
Foundation.
%
%

%
\end{document}